\documentstyle{article}
   \newcommand {\nc}{\newcommand}
   \nc{\eq}{\begin{equation}}
   \nc{\en}{\end{equation}}
   \nc{\eqa}{\begin{eqnarray}}
   \nc{\ena}{\end{eqnarray}}
   \nc{\eqann}{\begin{eqnarray*}}
   \nc{\enann}{\end{eqnarray*}}
   \nc {\dfn}[1]{{\it{#1}}}
   \nc{\nn}{\nonumber}
   \def\ep{\epsilon}
   
   \def\thv{theory }
   
   \def\gpv{group }

   \def\sbv{sub\gpv }

   \def\symgv{symmetry \gpv}

   \def\dgv{double \gpv}
   \def\rep{representation}
   \def\repv{representation }
   \def\Rep{Representation}
   \def\Repv{Representation }
   
   \def\repthv{\repv \thv}
   \def\indr{induced \rep}

   \def\svrep{single-valued \rep}
   
   \def\tvrep{two-valued \rep}
   
   \def\spinrep{spinor \rep}
   \def\spinrepv{spinor \repv}

   \def\irrv{irreducible }

   \def\alrep{inequivalent irreducible \rep s}
   \def\alrepv{inequivalent irreducible \rep s }
   
   \def\conv{conjugate }

   \def\dim{dimension}
   
   \def\dimv{dimension }
   \def\dimlv{\dim al }

   \def\clsf{classification}

   \nc{\sqt}{\sqrt{2}}
   \nc{\tsqt}{2\sqt}
   \nc{\msqt}{$\sqt$}
   \nc{\nsqt}{$-\sqt$}
   \nc{\mtsqt}{$\tsqt$}
   \nc{\ntsqt}{$-\tsqt$}

   \def\ot{\otimes}
   
   \def\smdp{>\hspace{-0.2cm}\lhd}

   \nc {\inv}[1]{#1^{-1}}
   \nc {\hc}[1]{{#1}^\dag}
   \nc {\cc}[1]{{#1}^\ast}
   \nc {\ad}[2]{Ad_{#1}({#2})}
   \nc {\wad}{\widetilde{Ad}}
   \nc {\wadf}[2]{\widetilde{Ad}_{#1}({#2})}
   \nc {\pb}[1]{{#1}^\ast}
   \def\tld{\tilde}
   \def\pr{\prime}

   \def\intg{{\cal Z}}
   \def\real{{\cal R}}
   \def\complex{{\cal C}}
   \def\quaternion{{\bf{H}}}
   \nc {\ga}[2]{{#1}[{#2}]}
   \nc {\cga}[1]{\ga{\complex}{#1}}
   \nc {\cgag}{\cga{G}}
   \nc {\eu}[1]{E^{#1}}
   \nc {\euf}{\eu{4}}
   \nc {\eun}{\eu{n}}
   \nc {\zn}[1]{\intg^{#1}}
   \nc {\zf}{\zn{4}}
   \nc {\zt} {Z_2}
   \nc {\ztn}[1]{\zt^{#1}}
   \nc {\ztt} {\ztn{2}}
   \nc {\ztth} {\ztn{3}}
   \nc {\ztf} {\ztn{4}}
   \nc {\ztnf}{\ztn{n}}
   \nc {\per}[1]{S_{#1}}
   \nc {\pern}{\per{n}}
   \nc {\irrr}[2]{IRR_{#1}({#2})}
   \nc {\irrc}[1]{\irrr{\complex}{#1}}
   \nc {\irrcg}{\irrc{G}}

   \nc {\mn}{(-)}
   \nc {\mns}[1]{\mn^{#1}}
   \nc {\mo}{(-1)}
   \nc {\mos}[1]{\mo^{#1}}
   \nc {\unit} {{\bf 1}}
   \nc {\unt} {\unit_{2\times 2}}
   \nc {\unth} {\unit_{3\times 3}}
   \nc {\unf} {\unit_{4\times 4}}
   \nc {\une} {\unit_{8\times 8}}
   \nc {\unw} {\unit_{12\times 12}}
   \nc {\zrt} {{\bf 0}_{2\times 2}}
   \nc {\zrth} {{\bf 0}_{3\times 3}}
   \nc {\zrf} {{\bf 0}_{4\times 4}}
    \def\CDalign#1{\bgroup\vcenter\bgroup\tabskip 2pt 
      \baselineskip 14pt \lineskip 3pt \lineskiplimit 3pt
      \halign\bgroup &\hfill$##$\hfill\crcr
      #1\crcr\egroup\egroup\egroup} 
    \def\CDto{\mathop{\relbar\joinrel\longrightarrow}\limits}    
    \def\CDup{\Big\uparrow}  
    \def\CDeq{\Big\|}
  \nc{\act}[3]{S_{{#1}}^{{#2}}[{#3}]}
  \nc{\cact}[2]{S_{Cl}^{{#1}}[{#2}]}
  
   \nc {\prj}[4]{\pi_{#1,#2}#2\ot_{S_#1}e^o_{#3,#4}}
   \nc {\prjf}[2]{\prj{o}{#1}{\eta}{#2}}
   \nc {\prjff}{\prjf{h}{i}}
   \nc {\Prj}[4]{\Pi_{#1,#2;#3,#4}}
   \nc {\Prjf}[2]{\Prj{o}{#1}{\eta}{#2}}
   \nc {\Prjff}{\Prjf{h}{i}}
   \nc {\orbt}[2]{\pi_{#1,#2}}
   \nc {\orbtf}[1]{\orbt{o}{#1}}
   \nc {\orbte}{\orbtf{e}}
   \nc {\rp}[3]{\orbt{#1}{#2}(#3)}
   \nc {\rpf}[2]{\rp{o}{#1}{#2}}
   \nc {\stbrp}[2]{D^o_#1(\tld{s}(#2))}
   \nc {\stbrpme}[4]{\stbrp{#1}{#2}^{#3}_{#4}}
   \nc {\stbrpf}[1]{\stbrp{\eta}{#1}}
   \nc {\stbrpmef}[3]{\stbrpme{\eta}{#1}{#2}{#3}}
   \nc {\ch}[2]{\chi_{#1;#2}}
   \nc {\che}[3]{\ch{#1}{#2}(#3)}
   \nc {\chf}[1]{\ch{o}{#1}}
   \nc {\chff}{\chf{\eta}}
   \nc {\chef}[2]{\che{o}{#1}{#2}}
   \nc {\cheff}[1]{\chef{\eta}{#1}}
   \nc {\chsb}[1]{\chi^o_{#1}}
   \nc {\chsbe}[2]{\chsb{#1}(#2)}
   \nc {\chsbf}{\chsb{\eta}}
   \nc {\chsbef}[1]{\chsbe{\eta}{#1}}
   \nc {\cg}[1]{O_{#1}}
   \nc {\cgn} {\cg{n}}
   \nc {\oh} {\cg{4}}
   \nc {\ohd} {{\overline{\oh}}}
   \nc {\ztfb} {\overline{\ztf}}
   \nc {\iso}[2]{ISO_{#1}(#2)}
   \nc {\isod}[1]{\iso{d}{#1}}
   \nc {\eg}[1]{ISO(#1)}
   \nc {\egn}{\eg{n}}
   \nc {\fs}[1]{F(#1)}
   \nc {\fsf}{\fs{\emb}}
   \nc {\fx}[1]{I(#1)}
   \nc {\fxf}{\fx{\emb}}
   \nc {\wrn}[1]{\ztn{#1}\smdp\per{#1}}
   \nc {\wn}{\wrn{n}}
   \nc {\wnn}{\pern^{\zt}}
   \nc {\w}[1]{\per{#1}^{\zt}}
   \nc {\fix}[1]{\per{(n-#1)}\ot \per{p}}
   \nc {\fixp}{\fix{p}}
   \nc {\cb}[1]{C_{#1}}
   \nc {\cn}{\cb{n}}

   \nc {\prm}{\sigma}
   \nc {\prmt}{\tld{\prm}}
   \nc {\prmp}{\prm^\pr}
   \nc {\cyc}[2]{\tau_{{#1}{#2}}}
   \nc {\emb}{\iota}
   \nc {\epy}{\tld{\ep}}
   \nc {\de}[1]{d_{E^{#1}}}
   \nc {\den}{\de{n}}
   \nc {\dy}{\tld{d}}
   \nc {\repw}[2]{e_{(#1)#2}}
   \nc {\repww}[4]{\repw{#1}{#2}\ot\repw{#3}{#4}}
   \nc {\prw}[6]{\pi_{#1,#2}#2\ot_{F_#1}(\repww{#3}{#4}{#5}{#6})}
   \nc {\dm}[1]{d_{({#1})}}
   \nc {\sls}{{\it slash}}
   \nc {\slsv}{{\it slash} }
   \nc {\lcl} {{\it local}}
   \nc {\lclv} {{\it local} }
   \nc {\drln}[5]
    {\put(#1,#2){\line(#3,#4){#5}}}
   \nc {\ybxa}[4]
    {
     \begin{picture}(40,10)
      \drln {0}{0}{0}{1}{10}
      \drln {0}{0}{1}{0}{40}
      \drln {10}{0}{0}{1}{10}
      \drln {0}{10}{1}{0}{40}
      \drln {20}{0}{0}{1}{10}
      \drln {30}{0}{0}{1}{10}
      \drln {40}{0}{0}{1}{10}
      \drln {0}{0}{1}{1}{#1}
      \drln {10}{0}{1}{1}{#2}
      \drln {20}{0}{1}{1}{#3}
      \drln {30}{0}{1}{1}{#4}
     \end{picture}
    }
   \nc {\ybxb}[4]
    {
     \begin{picture}(30,20)
      \put(0,0){\line(0,1){20}}
      \put(0,0){\line(1,0){10}}
      \put(10,0){\line(0,1){20}}
      \put(0,10){\line(1,0){30}}
      \put(0,20){\line(1,0){30}}
      \put(20,10){\line(0,1){10}}
      \put(30,10){\line(0,1){10}}
      \drln {0}{0}{1}{1}{#1}
      \drln {0}{10}{1}{1}{#2}
      \drln {10}{10}{1}{1}{#3}
      \drln {20}{10}{1}{1}{#4}
     \end{picture}
    }
   \nc {\ybxc}[4]
    {
     \begin{picture}(20,20)
      \put(0,0){\line(0,1){20}}
      \put(0,0){\line(1,0){20}}
      \put(10,0){\line(0,1){20}}
      \put(0,10){\line(1,0){20}}
      \put(0,20){\line(1,0){20}}
      \put(20,0){\line(0,1){20}}
      \drln {0}{0}{1}{1}{#1}
      \drln {10}{0}{1}{1}{#2}
      \drln {0}{10}{1}{1}{#3}
      \drln {10}{10}{1}{1}{#4}
     \end{picture}
    }
   \nc {\ybxd}[4]
    {
     \begin{picture}(20,30)
      \drln {0}{0}{0}{1}{30}
      \drln {0}{0}{1}{0}{10}
      \drln {10}{0}{0}{1}{30}
      \drln {0}{10}{1}{0}{10}
      \drln {0}{20}{1}{0}{20}
      \drln {0}{30}{1}{0}{20}
      \drln {20}{20}{0}{1}{10}
      \drln {0}{0}{1}{1}{#1}
      \drln {0}{10}{1}{1}{#2}
      \drln {0}{20}{1}{1}{#3}
      \drln {10}{20}{1}{1}{#4}
     \end{picture}
    }
   \nc {\ybxe}[4]
    {
     \begin{picture}(10,40)
      \drln {0}{0}{0}{1}{40}
      \drln {10}{0}{0}{1}{40}
      \drln {0}{0}{1}{0}{10}
      \drln {0}{10}{1}{0}{10}
      \drln {0}{20}{1}{0}{10}
      \drln {0}{30}{1}{0}{10}
      \drln {0}{40}{1}{0}{10}
      \drln {0}{0}{1}{1}{#1}
      \drln {0}{10}{1}{1}{#2}
      \drln {0}{20}{1}{1}{#3}
      \drln {0}{30}{1}{1}{#4}
     \end{picture}
    }
   
   \nc{\op}{orientation-preserved}
   \nc{\opv}{orientation-preserved }
   \nc{\on}[1]{SO_{#1}}
   \nc{\ohn}[1]{O_{#1}}
   \nc{\of}{\on{4}}
   \nc{\ohf}{\ohn{4}}
   \nc{\ofd}{\overline{\of}}
   \nc{\ohfd}{\overline{\ohf}}
   \nc{\cir}[1]{e^{i({#1})\pi}}
 \title{Sinle-valued and \spinrep s of orientation-preserved four-\dimlv cubic group}
 \author{Jian Dai\thanks{daij1492@yahoo.com},
  Xing-Chang Song\thanks{songxc@ibm320h.phy.pku.edu.cn}\\
  Department of Physics, Peking University}
 \date{February 23, 1999\\
   Revised on September 6, 2000}
 
\begin{document}
  \begin{titlepage}
  \maketitle
  \begin{abstract}
  \noindent
   We calculate the single-valued and spinor \rep s of $\of$, the \opv \sbv of
   $\ohf$, on the base of our previous work on four-dimensional cubic group $\ohf$,
  \end{abstract}
  \end{titlepage}
  \section{Introduction}
   This paper should be considered as a continuation of our
   previous work \cite{ds} where the \svrep s and \spinrep s of four-\dimlv cubic group
   $\ohf$ are obtained. We will consider here the orientation-preserved \sbv of $\ohf$, namely
   $\of$, and derive the \repthv of its \dgv $\ofd$. The \repv theories of $\of$ and $\ohf$ are well-known results
   for years \cite{baake}\cite{mandula}; we re-derive them adopting a
   more systematic method and give their spinor parts
   which are totally new. This paper is arranged as following. First we construct $\of$, extend it to double group $\ofd$ and calculate their
   \conv classes. Then we reduce all \svrep s as well as \tvrep s of $\of$ from those of
   $\ohf$ based on our previous work. From the point of physical application,
   we expect that fermions in lattice field theory should
   transform according to \spinrep s of \symgv of the lattice where they
   reside.
  \section{Structure of $\ofd$}
   For completeness, we recall some definitions and results in \cite{ds}.
   An \dfn{$n$-cube} (or \dfn{hyper-cube in $\eun$}) $\cn$ is defined to
   be a subset of $\eun$, $\cn=\{p|x^i(p)=\pm 1\}$, where $x^i:\eun\rightarrow \real, i=1,2,...,n$
   are coordinate functions of $\eun$, together with the distance inherited from $\eun$.
   \dfn{$n$-Cubic group} (\dfn{hyper-cubic group of degree $n$}) $O_n$ consists of all isometries of $\eun$ which stabilize
   $\cn$ and has a structure of wreath product
   \eq
   \label{wreath}
    \cgn\cong \ztnf\smdp \pern
   \en
   in which the first factor $\ztn{n}$ is generated by the inversions of axes
   and the second factor $S_n$ is the permutations
   of axes. The wreath product structure (\ref{wreath}) enable us to regard $\cgn$ as a permutation group with signatures whose
   elements can be decomposed into independent cycles with
   signature. A $k$-cycle with signature in $\wn$ is represented
   as
   \[
    \left(
       \begin{array}{cccc}
        a_1              &a_2              &\ldots &a_k            \\
        \mns{s_{a_1}}a_2 &\mns{s_{a_2}}a_3 &\ldots &\mns{s_{a_k}}a_1
       \end{array}
      \right), s_{a_i}\in\intg/2\intg, i=1,2,...,k
   \]
   whose signature is defined to be
   $\mns{s_{a_1}+s_{a_2}+...+s_{a_k}}$. A \conv classes of $\cgn$ can be uniquely
   determined by the cycle structure descended from $\pern$ and signatures of its cycles \cite{kerber}. \\
   \\
   Let $n=4$ below. $\ohf$ is generated by a set of generators $\{e_i, i=1,2,3,4; \gamma, t\}$ together with a system of
   constraint equations
   \eqa
   \label{o4g1}
    e_i^2=e, e_ie_j=e_je_i, i,j=1..4, i\not= j\\
   \label{o4g2}
    \gamma^2=e, t^3=e,(t\gamma)^4=e\\
   \label{o4g3}
    t e_1=e_1,t e_2=e_4 t,t e_3=e_2 t, t e_4=e_3 t\\
   \label{o4g4}
    \gamma e_1=e_3\gamma,\gamma e_2=e_2\gamma,\gamma e_4=e_4\gamma
   \ena
   in which $e_i$s generate $\ztn{4}$ and $\gamma$, $t$ generate $S_4$.
   All of 20 \conv classes of $\ohf$ can be depicted by a class of generalized Young
   diagrams where a slash inside a Young box shows that there is a minus sign in the position of corresponding cycle
   (see table \ref{tab1}).
   \begin{table}[d]
    \caption{Conjugate Classes of $\ztn{4}\smdp\per{4}$}
    \label{tab1}
     \begin {tabbing}
      xxx\=xxxxxxxx\=xxxxxxxxxxxxxxxxx\=xxx\=xxxxx\=xxxxxxxxxx\=xxx\=xxxxxxxx\=xxxxxxxxxxxxxxxxx\=xxx\=xxxxx\=xxx \kill
      No\>SplitNo\>YoungDiagram\>ord\>num\>det \>No\>SplitNo\>YoungDiagram\>ord\>num\>det\\
      1\>1-1\>\ybxa{0}{0}{0}{0}\>1\>1\>1\>2\>1-2\>\ybxa{0}{0}{0}{10}\>2\>4\>-1\\
      3\>1-3\>\ybxa{0}{0}{10}{10}\>2\>6\>1\>4\>1-4\>\ybxa{0}{10}{10}{10}\>2\>4\>-1\\
      5\>1-5\>\ybxa{10}{10}{10}{10}\>2\>1\>1\\
      6\>2-1\>\ybxb{0}{0}{0}{0}\>2\>12\>-1\>7\>2-2\>\ybxb{0}{0}{0}{10}\>2\>24\>1\\
      8\>2-3\>\ybxb{10}{0}{0}{0}\>4\>12\>1\>9\>2-4\>\ybxb{0}{0}{10}{10}\>2\>12\>-1\\
      10\>2-5\>\ybxb{10}{0}{0}{10}\>4\>24\>-1\>11\>2-6\>\ybxb{10}{0}{10}{10}\>4\>12\>1\\
      12\>3-1\>\ybxc{0}{0}{0}{0}\>2\>12\>1\>13\>3-2\>\ybxc{0}{10}{0}{0}\>4\>24\>-1\\
      14\>3-3\>\ybxc{10}{10}{0}{0}\>4\>12\>1\\
      15\>4-1\>\ybxd{0}{0}{0}{0}\>3\>32\>1\>16\>4-2\>\ybxd{0}{0}{0}{10}\>6\>32\>-1\\
      17\>4-3\>\ybxd{10}{0}{0}{0}\>6\>32\>-1\>18\>4-4\>\ybxd{10}{0}{0}{10}\>6\>32\>1 \\
      19\>5-1\>\ybxe{0}{0}{0}{0}\>4\>48\>-1\>20\>5-2\>\ybxe{10}{0}{0}{0}\>8\>48\>1  \\
     \end{tabbing}
     "SplitNo" of each class reflects the relation between the
     classes of $\ztn{4}\smdp\per{4}$ and those of $\per{4}$. "ord" means order of each class.
     "num" is the number of elements in each class. "det" is the
     signature of each class.
    \end{table}
    \\
    \\
   We know from the theory of Clifford algebra \cite{harvey} that there is a short exact
   sequence
   \[
    0\rightarrow Z_2\rightarrow
    Pin(E^n)\stackrel{\pi}{\rightarrow} O(n)\rightarrow 1
   \]
   Let $\ep$ be an injective homomorphism from a \gpv $G$ to $O(n)$, then the double group or the spin-extension of $G$
   with respect to $\ep$ is defined to be $D_n(G,\ep):=\pi^{-1}(G)$.
   With $n$ and $\epsilon$ fixed, we use $\bar{G}$ to denote
   $D_n(G, \epsilon)$. Let $C$ be a \conv class in $G$, then either will $\pi^{-1}(C)$
   be one \conv class in $\bar{G}$ still or it will split into two \conv classes in
   $\bar{G}$ containing $s(g), -s(g), \forall g\in C$ respetively, where $s:
   G\rightarrow \bar{G}, s.t. \pi s=Id_G$.
   \\
   \\
   $\ohfd$ is generated by lifting
   Eq.(\ref{o4g1})(\ref{o4g2})(\ref{o4g3})(\ref{o4g4}) to
   \eqa
   \label{o4bg1}
    e_i^2=-1, e_ie_j=-e_je_i, i,j=1..4, i\not= j\\
   \label{o4bg2}
    \gamma^2=-1, t^3=-1, (t\gamma)^4=-1\\
   \label{o4bg3}
    te_1=e_1t, te_2=-e_4t, te_3=-e_2t, te_4=e_3 t\\
   \label{o4bg4}
    \gamma e_1=-e_3\gamma, \gamma e_2=-e_2\gamma, \gamma e_4=-e_4\gamma
   \ena
   Each \conv class of $\ohf$ is lifted into one corresponding \conv class in $\ohfd$, except
   classes $1,8,14,15,20$ which splits into two classes each of them. Hence there are total 25 classes of $\ohfd$.\\
   \\
   The {\it orientation-preserved n-cubic group} is defined to be
   \[
    SO_n:=(\cgn\cap SO(n))\lhd\cgn
   \]
   The generators of $\ztn{n}$ are written as $e_i, i=1,2,...,n$. Define
   $\ztn{n}|_e$ as a \sbv of $\ztn{n}$ generated by $e_ie_j, i\neq
   j$ and $\ztn{n}|_o:=\ztn{n}\backslash \ztn{n}|_e$.
   Then
    \eq
    \label{decomOn}
     \on{n}=(\ztn{n}|_e\cdot A_{n})\bigsqcup (\ztn{n}|_o\cdot (S_{n}\backslash A_{n}))
    \en
    in which $\cdot$ is product of two subsets in a group, $A_{n}$
    stands for the alternative \sbv in $S_{n}$.
   Thus, $|\on{n}|=(2^{n}\cdot (n)!)/2$.\\
   \\
   Specify $n=4$ and we know immediately that $|\of|=192$.
   We introduce
   \eqa
   \label{def1}
    \eta=\gamma t\gamma t^2\gamma;\\
   \label{def2}
    \alpha=(t^2\gamma)^2, \beta=t\gamma t^2\gamma t;\\
   \label{def3}
    x=e_1\eta, y= e_4\eta, q=e_2\eta
   \ena
   then the structure of $\of$ can be summarized as
   \eqa
   \label{so4g1}
    x^2=y^2=q^4=e,yx=xy,qx=xq^3,qy=yq^3\\
   \label{so4g2}
    \alpha^2=\beta^2=t^3=e,
    \beta\alpha=\alpha\beta,
    t\alpha=\alpha\beta t, t\beta=\alpha t\\
   \label{so4g3}
    \alpha x=q\beta,\alpha y=q^3\beta,\alpha q=x\beta\\
   \label{so4g4}
    \beta x=q^3\alpha, \beta y=q\alpha,\beta q=y\alpha\\
   \label{so4g5}
    tx=xt^2,ty=q^3t^2,tq=yt^2
   \ena
   together with
   \eq
   \label{so4g6}
    x\alpha=\beta q^3, xt=t^2x; y\beta=\alpha q^3, yt=t^2q^3;
    qt=t^2 y
   \en
   Accordingly, each group element can be expressed as a "normal ordering" product of $x,y\rightarrow q\rightarrow
   \alpha,\beta\rightarrow t$ and their powers from left to right.
   Throwing away all classes which belong to $\ohf$ but not to $\of$,
   there are 11 left which are $1,3,5,7,8,11,12,14,15,18,20$ in Table \ref{tab1}. The $14$th and the $20$th will part
   into two classes with equal numbers of elements each under adjoint action of
   $\of$ which are denoted as $14, 14^\pr, 20, 20^\pr$.
   Therefore, there are 13 \conv classes in $\of$.\\
   \\
   Due to the fact that $\ofd\lhd \ohfd$ and theory of group extension \cite{brown},
   the diagram
   \[
    \CDalign{0&\CDto&Z_2  &\CDto &\ohfd &\CDto^{\pi}&\ohf &\CDto&1 \cr
              &     &\CDeq&      &\CDup &           &\CDup&        \cr
             0&\CDto&Z_2  &\CDto &\ofd  &\CDto      &\of  &\CDto&1  }
   \]
   is commutative. We can lift generating relations
   (\ref{so4g1})(\ref{so4g2})...(\ref{so4g6})
   to
   \eqann
    x^2=y^2=q^4=-1,yx=-xy,qx=-xq^3,qy=-yq^3\\
    \alpha^2=\beta^2=t^3=-1,
    \beta\alpha=-\alpha\beta,
    t\alpha=-\alpha\beta t, t\beta=\alpha t\\
    \alpha x=q\beta,\alpha y=q^3\beta,\alpha q=x\beta\\
    \beta x=q^3\alpha, \beta y=q\alpha,\beta q=y\alpha\\
    tx=-xt^2,ty=q^3t^2,tq=yt^2\\
    x\alpha=\beta q^3, xt=-t^2x; y\beta=\alpha q^3, yt=t^2q^3;
    qt=t^2 y
   \enann
   by the definition (\ref{def1})(\ref{def2})(\ref{def3}).
   As to the \conv \clsf, $1,5,8,11,14,15,18,20,14^\pr,20^\pr$ violate relation $-g\sim g$,
   so that $\ofd$ is partitioned into 23 classes, suggesting that there are altogether 23
   \alrepv in which 13 \rep s are single-valued to $\of$.
   We summarize the \conv classes of $\ofd$ in Table \ref{tab3}.\\
   \footnotesize
   \begin{table}[h]
   \caption{Conjugate Classes of $\ofd$}
   \label{tab3}
   \begin{tabular}{|l|r|r|r|r|r|r|r|r|r|r|r|r|r|r|r|r|r|r|r|r|r|r|r|}\hline
     No.&$1$&$\bar{1}$&3&$5$&$\bar{5}$&7&$8$&$\bar{8}$&$11$&$\bar{11}$&12&$14$&$\bar{14}$&$14^\pr$&$\bar{14^\pr}$
     &$15$&$\bar{15}$&$18$&$\bar{18}$&$20$&$\bar{20}$&$20^\pr$&$\bar{20^\pr}$\\ \hline
     num.&1&1&12&1&1&48&12&12&12&12&24&6&6&6&6&32&32&32&32&24&24&24&24\\ \hline
     ord.&1&2&4&2&2&4&8&8&8&8&4&4&4&4&4&6&3&6&6&8&8&8&8\\ \hline
   \end{tabular}
   \end{table}
   \normalsize
  \section{\Repv theory of $\ofd$}
   All \alrepv of $\ofd$ can be reduced out from those of $\ohfd$~\cite{ds}.
   Table \ref{tab4} gives the characters of all \alrepv of $\ohfd$, respect to the classes of
   $\ofd$.
   \begin{table}[h]
   \caption{Character Table of $\ohfd$ (with respect to the classes of $\ofd$)}
   \label{tab4}
   \tabcolsep 2pt
   \small{
   \begin{tabular}{|l|c|c|c|c|c|c|c|c|c|c|c|c|c|c|c|c|c|c|c|c|c|c|c|c|}\hline
    &1&$\bar{1}$&3&5&$\bar{5}$&7&8&$\bar{8}$&11&$\bar{11}$&12&14&$\bar{14}$
    &$14^\pr$&$\bar{14^\pr}$&15&$\bar{15}$&18&$\bar{18}$&20&$\bar{20}$&$20^\pr$
    &$\bar{20^\pr}$&$(\chi,\chi)$\\\hline
   $\chi_1^{(1)}$&1&1&1&1&1&1&1&1&1&1&1&1&1&1&1&1&1&1&1&1&1&1&1&{\bf 1}\\\hline
   $\chi_2^{(1)}$&1&1&1&1&1&$-1$&$-1$&$-1$&$-1$&$-1$
    &1&1&1&1&1&1&1&1&1&$-1$&$-1$&$-1$&$-1$&{\bf 1}\\\hline
   $\chi_1^{(2)}$&2&2&2&2&2&0&0&0&0&0&2&2&2&2&2&$-1$&$-1$&$-1$&$-1$&0&0&0&0&{\bf 1}\\\hline
   $\chi_1^{(3)}$&3&3&3&3&3&1&1&1&1&1&$-1$&$-1$&$-1$&$-1$&$-1$
    &0&0&0&0&$-1$&$-1$&$-1$&$-1$&{\bf 1}\\\hline
   $\chi_2^{(3)}$&3&3&3&3&3&$-1$&$-1$&$-1$&$-1$&$-1$&$-1$&$-1$&$-1$&$-1$&$-1$
    &0&0&0&0&1&1&1&1&{\bf 1}\\\hline
   $\chi_3^{(1)}$&1&1&1&1&1&$-1$&$-1$&$-1$&$-1$&$-1$
    &1&1&1&1&1&1&1&1&1&$-1$&$-1$&$-1$&$-1$&{\bf 1}\\\hline
   $\chi_4^{(1)}$&1&1&1&1&1&1&1&1&1&1&1&1&1&1&1&1&1&1&1&1&1&1&1&{\bf 1}\\\hline
   $\chi_2^{(2)}$&2&2&2&2&2&0&0&0&0&0&2&2&2&2&2&$-1$&$-1$&$-1$&$-1$&0&0&0&0&{\bf 1}\\\hline
   $\chi_3^{(3)}$&3&3&3&3&3&$-1$&$-1$&$-1$&$-1$&$-1$&$-1$&$-1$&$-1$&$-1$&$-1$
    &0&0&0&0&1&1&1&1&{\bf 1}\\\hline
   $\chi_4^{(3)}$&3&3&3&3&3&1&1&1&1&1&$-1$&$-1$&$-1$&$-1$&$-1$
    &0&0&0&0&$-1$&$-1$&$-1$&$-1$&{\bf 1}\\\hline
   $\chi_1^{(4)}$&4&4&0&$-4$&$-4$&0&2&2&$-2$&$-2$&0&0&0&0&0&1&1&$-1$&$-1$
    &0&0&0&0&{\bf 1}\\\hline
   $\chi_2^{(4)}$&4&4&0&$-4$&$-4$&0&$-2$&$-2$&2&2&0&0&0&0&0&1&1&$-1$&$-1$
    &0&0&0&0&{\bf 1}\\\hline
   $\chi_1^{(8)}$&8&8&0&$-8$&$-8$&0&0&0&0&0&0&0&0&0&0&$-1$&$-1$&1&1
    &0&0&0&0&{\bf 1}\\\hline
   $\chi_3^{(4)}$&4&4&0&$-4$&$-4$&0&$-2$&$-2$&2&2&0&0&0&0&0&1&1&$-1$&$-1$
    &0&0&0&0&{\bf 1}\\\hline
   $\chi_4^{(4)}$&4&4&0&$-4$&$-4$&0&2&2&$-2$&$-2$&0&0&0&0&0&1&1&$-1$&$-1$
    &0&0&0&0&{\bf 1}\\\hline
   $\chi_2^{(8)}$&8&8&0&$-8$&$-8$&0&0&0&0&0&0&0&0&0&0&$-1$&$-1$&1&1
    &0&0&0&0&{\bf 1}\\\hline
   $\chi_1^{(6)}$&6&6&$-2$&6&6&0&0&0&0&0&2&$-2$&$-2$&$-2$&$-2$&0&0&0&0
    &0&0&0&0&{\bf 1}\\\hline
   $\chi_2^{(6)}$&6&6&$-2$&6&6&2&-2&-2&-2&-2&-2&2&2&2&2&0&0&0&0
    &0&0&0&0&{\bf 2}\\\hline
   $\chi_3^{(6)}$&6&6&$-2$&6&6&0&0&0&0&0&2&$-2$&$-2$&$-2$&$-2$&0&0&0&0
    &0&0&0&0&{\bf 1}\\\hline
   $\chi_4^{(6)}$&6&6&$-2$&6&6&-2&2&2&2&2&-2&2&2&2&2&0&0&0&0
    &0&0&0&0&{\bf 2}\\\hline
   $\chi_1^{(\b{4})}$&4&-4&0&0&0&0&\ntsqt&\mtsqt&0&0&0&2&-2&2&-2
    &2&-2&0&0&\msqt&\nsqt&\msqt&\nsqt&{\bf 2}\\\hline
   $\chi_2^{(\b{4})}$&4&-4&0&0&0&0&\mtsqt&\ntsqt&0&0&0&2&-2&2&-2
    &2&-2&0&0&\nsqt&\msqt&\nsqt&\msqt&{\bf 2}\\\hline
   $\chi^{(\b{8})}$&8&-8&0&0&0&0&0&0&0&0&0&4&-4&4&-4
    &-2&2&0&0&0&0&0&0&{\bf 2}\\\hline
   $\chi_1^{(\b{12})}$&12&-12&0&0&0&0&\ntsqt&\mtsqt&0&0&0&-2&2&-2&2
    &0&0&0&0&\nsqt&\msqt&\nsqt&\msqt&{\bf 2}\\\hline
   $\chi_2^{(\b{12})}$&12&-12&0&0&0&0&\mtsqt&\ntsqt&0&0&0&-2&2&-2&2
    &0&0&0&0&\msqt&\nsqt&\msqt&\nsqt&{\bf 2}\\\hline
   \end{tabular}}
   \normalsize\\
   \\
   Character is labeled by a superscript showing its \dimv where an underline shows a \spinrepv
   and a subscript distinguishing different \rep s with same \dim. $(\chi,\chi)$ evaluates the inner product of a character with
   itself.
   \end{table}
   A brief observation gives some important information. Firstly, $1_1\cong 1_4,1_2\cong 1_3,2_1\cong 2_2,3_1\cong 3_4,
   3_2\cong 3_3, 4_1\cong 4_4,4_2\cong 4_3,8_1\cong 8_2,6_1\cong 6_3$. Secondly,
   omitting equivalence, $1_1$, $1_2$, $2_1$, $3_1$, $3_1$, $4_1$, $4_2$, $8_1$, $6_1$,
   remain irreducible within $\ofd$, while other 7 become reducible. Thirdly, as for each of
   these reducible ones, the inner product of itself's character equals to 2, implying
   that it can be reduced to two \alrepv, thus there are 13 single-valued and 10
   spinor \rep s as we expected. Finally, it is one possible solution to Burside theorem
   that each of these seven reducible \rep s splits into 2 \alrepv
   with equal dimensions. We conjecture that it is the solution to our \repv
   theory of $\ofd$ and try to verify it below.
   \\
   \\
   Summarily speaking, there are nine single-valued \alrep s inherited from $\ohfd$
   \[
    1_1,1_2,2\equiv 2_1,3_1,3_2,4_1,4_2,8\equiv 8_1,6\equiv 6_1
   \]
   and we conjecture the splitting relations
   \eqann
    6_2, 6_4\rightarrow 3_{\alpha}, 3_{\beta}, 3_{\gamma}, 3_{\delta}\\
    \b{4}_1, \b{4}_2\rightarrow 2_{\alpha}, 2_{\beta}, 2_{\gamma},2_{\delta}\\
    \b{8}\rightarrow 4_{\alpha}, 4_{\beta}\\
    \b{12}_1, \b{12}_2\rightarrow 6_{\alpha}, 6_{\beta}, 6_{\gamma}, 6_{\delta}
   \enann
   \subsection{Hidden single-valued \rep s}
    The \repv matrices of $x$, $y$, $q$ in $6_2$~\cite{ds} are written as
    \[
     x\mapsto
      \left(
       \begin{array}{cccccc}
        1&0&0&0&0&0\\
        0&1&0&0&0&0\\
        0&0&0&1&0&0\\
        0&0&1&0&0&0\\
        0&0&0&0&0&-1\\
        0&0&0&0&-1&0
       \end{array}
      \right),
     y\mapsto
      \left(
       \begin{array}{cccccc}
        1&0&0&0&0&0\\
        0&1&0&0&0&0\\
        0&0&0&-1&0&0\\
        0&0&-1&0&0&0\\
        0&0&0&0&0&1\\
        0&0&0&0&1&0
       \end{array}
      \right),
     q\mapsto
      \left(
       \begin{array}{cccccc}
        -1&0&0&0&0&0\\
        0&-1&0&0&0&0\\
        0&0&0&-1&0&0\\
        0&0&1&0&0&0\\
        0&0&0&0&0&1\\
        0&0&0&0&-1&0
       \end{array}
      \right)
   \]
    The textures of these matrices inspire us to such a hypotheses that in
    $3_{\alpha,\beta,\gamma,\delta}$, x,y,q take on a form like
    \[
      x,y\mapsto
       \left(
        \begin{array}{cc}
         \pm 1&\\&\pm H
        \end{array}
       \right),
      q\mapsto
       \left(
        \begin{array}{cc}
         \pm 1&\\&\pm Q
        \end{array}
       \right)
    \]
    where $H\equiv
       \left(
        \begin{array}{cc}
         0&1\\1&0
        \end{array}
       \right),
      Q\equiv
       \left(
        \begin{array}{cc}
         0&-1\\1&0
        \end{array}
       \right)$.
    After taking account of the \conv equivalence, only four possibilities survive
    from the totally 64, i.e.
    \eq
    \label{I}
     I:
       x\rightarrow
        \left(
         \begin{array}{cc}
          1&\\&H
         \end{array}
        \right),
       y\rightarrow
        \left(
         \begin{array}{cc}
          1&\\&-H
         \end{array}
        \right),
       q\rightarrow
        \left(
         \begin{array}{cc}
          -1&\\&Q
         \end{array}
        \right)
    \en
    \eq
    \label{II}
     II:
       x\rightarrow
        \left(
         \begin{array}{cc}
          -1&\\&H
         \end{array}
        \right),
       y\rightarrow
        \left(
         \begin{array}{cc}
          -1&\\&-H
         \end{array}
        \right),
       q\rightarrow
        \left(
         \begin{array}{cc}
          1&\\&Q
         \end{array}
        \right)
    \en
    \eq
    \label{III}
     III:
       x\rightarrow
        \left(
         \begin{array}{cc}
          1&\\&-H
         \end{array}
        \right),
       y\rightarrow
        \left(
         \begin{array}{cc}
          1&\\&H
         \end{array}
        \right),
       q\rightarrow
        \left(
         \begin{array}{cc}
          -1&\\&Q
         \end{array}
        \right)
    \en
    \eq
    \label{IV}
     IV:
       x\rightarrow
        \left(
         \begin{array}{cc}
          -1&\\&-H
         \end{array}
        \right),
       y\rightarrow
        \left(
         \begin{array}{cc}
          -1&\\&H
         \end{array}
        \right),
       q\rightarrow
        \left(
         \begin{array}{cc}
          1&\\&Q
         \end{array}
        \right)
    \en
    which also satisfy Eq.(\ref{so4g1}).
    Then we regard $\alpha,\beta,t$ as unknowns, (\ref{so4g2})..(\ref{so4g6}) as constraint,
    and solve these matrix equations. Modulo similarity, each of Eqs.(\ref{I})(\ref{II})
    (\ref{III})(\ref{IV}) gives two solutions,
    labeled as $I$,$I'$,$II$,$II'$,$III$,$III'$,$IV$,$IV'$;
    however, there is no difficulty to find out that $I\cong III$, $I'\cong III'$,
    $II\cong IV$, $II'\cong IV'$. Thus $I$,$I'$,$II$,$II'$ are what we need.
    \eqann
     3_{\alpha}\equiv I:
      \alpha\rightarrow diag(-1,1,-1),
      \beta\rightarrow diag(1,-1,-1),
      t\rightarrow
       \left(
        \begin{array}{ccc}
         0&0&1\\1&0&0\\0&1&0
        \end{array}
       \right)\\
     3_{\beta}\equiv I':
      \alpha\rightarrow diag(1,-1,-1),
      \beta\rightarrow diag(-1,-1,1),
      t\rightarrow
       \left(
        \begin{array}{ccc}
         0&0&1\\1&0&0\\0&1&0
        \end{array}
       \right)\\
     3_{\gamma}\equiv II:
      \alpha\rightarrow diag(-1,1,-1),
      \beta\rightarrow diag(1,-1,-1),
      t\rightarrow
       \left(
        \begin{array}{ccc}
         0&0&1\\-1&0&0\\0&-1&0
        \end{array}
       \right)\\
     3_{\delta}\equiv II':
      \alpha\rightarrow diag(1,-1,-1),
      \beta\rightarrow diag(-1,-1,1),
      t\rightarrow
       \left(
        \begin{array}{ccc}
         0&0&1\\-1&0&0\\0&-1&0
        \end{array}
       \right)
    \enann
   \subsection{Spinor \rep s}
    It is more straightforward to reduce out the spinor \rep s.
    A deep result of Clifford theory
    on decomposition of \indr s \cite{cr}\cite{ds} shows
    that the spinor \repv matrices of
    $\ohf$ are of the form of a tensor product
    \[
     S_i(g)=S(g)\otimes s_i(g),
     \forall g\in\ohfd, i=\b{4}_1,\b{4}_2,\b{8},\b{12}_1,\b{12}_2
    \]
    where $S$ is given by the algebraic isomorphism from
    $Cl(E^4)$ to $M_2(\quaternion)$ and $s_i$ has the same texture (zero matrix element positions) of
    \irrv \repv $i$ of $\per{4}$.
    Additionally, for $g$ in $\ofd$, $S(g)$ takes on a 2-by-2
    block diagonal form
    \eq
     S(g)=\left(
       \begin{array}{cc}
        S_{up}(g)&0\\0&S_{down}(g)
       \end{array}
      \right)
    \en
    So it is just what we want
    \eqa
     S_{up}(x)= {1\over\sqt}\cdot
       \left(
        \begin{array}{cc}
         \cir{3\over 4}&\cir{-3\over 4}\\ \cir{1\over 4}&\cir{-1\over 4}
        \end{array}
       \right),
     S_{up}(y)={1\over\sqt}\cdot
       \left(
        \begin{array}{cc}
         \cir{3\over 4}&\cir{1\over 4}\\ \cir{-3\over 4}&\cir{-1\over 4}
        \end{array}
       \right)\\
      S_{up}(q)={1\over\sqt}\cdot
       \left(
        \begin{array}{cc}
         \cir{1\over 4}&\cir{3\over 4}\\ \cir{3\over 4}&\cir{1\over 4}
        \end{array}
       \right),
      S_{up}(t)={1\over\sqt}\cdot
       \left(
        \begin{array}{cc}
         1&-i\\1&i
        \end{array}
       \right)\\
      S_{up}(\alpha)=
       \left(
        \begin{array}{cc}
         0&-i\\-i&0
        \end{array}
       \right),
      S_{up}(\beta)=
       \left(
        \begin{array}{cc}
         -i&0\\0&i
        \end{array}
       \right)\\
     S_{down}(x)={1\over\sqt}\cdot
       \left(
        \begin{array}{ccc}
         \cir{3\over 4}&\cir{-3\over 4}\\ \cir{1\over 4}&\cir{-1\over 4}
        \end{array}
       \right),
     S_{down}(y)\rightarrow {1\over\sqt}\cdot
       \left(
        \begin{array}{ccc}
         \cir{-1\over 4}&\cir{-3\over 4}\\ \cir{1\over 4}&\cir{3\over 4}
        \end{array}
       \right)\\
     S_{down}(q)={1\over\sqt}\cdot
       \left(
        \begin{array}{ccc}
         \cir{1\over 4}&\cir{3\over 4}\\ \cir{3\over 4}&\cir{1\over 4}
        \end{array}
       \right),
     S_{down}(t)={1\over\sqt}\cdot
       \left(
        \begin{array}{cc}
         i&1\\-i&1
        \end{array}
       \right)\\
     S_{down}(\alpha)=
       \left(
        \begin{array}{cc}
         0&1\\-1&0
        \end{array}
       \right),
     S_{down}(\beta)=
       \left(
        \begin{array}{cc}
         0&-i\\-i&0
        \end{array}
       \right)
    \ena
    Keeping the second factor
    unchanged, each spinor \repv in $\ohfd$ splits into two spinor \rep s in $\ofd$,
    denoted as $2_\alpha$,$2_\beta$,$2_\gamma$,$2_\delta$,$4_\alpha$,$4_\beta$,
    $6_\alpha$,$6_\beta$,$6_\gamma$,$6_\delta$.\\
    \\
    In fact, $S(g)$ falls in the
    so-called "chiral"-\repv of $Cl(E^4)$ in physical language.
    Due to $Cl(V)=Cl(V)_e\oplus Cl(V)_o$, and the choice of chiral-\rep, there are
    \[
     Cl(E^4)_e\cong
       \left(
       \begin{array}{cc}
        {\bf H}&O\\0&{\bf H}
       \end{array}
      \right),
     Cl(E^4)_o\cong
       \left(
       \begin{array}{cc}
        0&{\bf H}\\{\bf H}&0
       \end{array}
      \right)
    \]
    Notice that $\ofd<Spin(4)\subset Cl(E^4)_e$, so our
    reducing process for spinor \rep s roots in the structure of Clifford algebra.
   \subsection{Results}
   Conclusively, our conjecture really gives all \alrep s whose characters are summarized in Table \ref{tab5}.
   \\
   \footnotesize
   \tabcolsep 3pt
   \begin{table}[h]
   \caption{Character Table of $\ofd$}
   \label{tab5}
    \begin{tabular}{|l||c|c|c|c|c|c|c|c|c|c|c|c|c|c|c|c|c|c|c|c|c|c|c|}\hline
     &$1$&$\bar{1}$&$3$&$5$&$\bar{5}$&$7$&8&$\bar{8}$&$11$&$\bar{11}$&$12$&$14$&$\bar{14}$
     &$14^{'}$&$\bar{14^{'}}$&$15$&$\bar{15}$&$18$&$\bar{18}$&$20$&$\bar{20}$&$20^{'}$
     &$\bar{20^{'}}$\\\hline
     num.&1&1&12&1&1&48&12&12&12&12&24&6&6&6&6&32&32&32&32&24&24&24&24\\
     \hline\hline
     $\chi_1^{(1)}$&1&1&1&1&1&1&1&1&1&1&1&1&1&1&1&1&1&1&1&1&1&1&1\\\hline
     $\chi_2^{(1)}$&1&1&1&1&1&$-1$&$-1$&$-1$&$-1$&$-1$
      &1&1&1&1&1&1&1&1&1&$-1$&$-1$&$-1$&$-1$\\\hline
     $\chi^{(2)}$&2&2&2&2&2&0&0&0&0&0&2&2&2&2&2&$-1$&$-1$&$-1$&$-1$&0&0&0&0\\\hline
     $\chi_1^{(3)}$&3&3&3&3&3&1&1&1&1&1&$-1$&$-1$&$-1$&$-1$&$-1$
      &0&0&0&0&$-1$&$-1$&$-1$&$-1$\\\hline
     $\chi_2^{(3)}$&3&3&3&3&3&$-1$&$-1$&$-1$&$-1$&$-1$&$-1$&$-1$&$-1$&$-1$&$-1$
      &0&0&0&0&1&1&1&1\\\hline
     $\chi_1^{(4)}$&4&4&0&$-4$&$-4$&0&2&2&$-2$&$-2$&0&0&0&0&0&1&1&$-1$&$-1$
      &0&0&0&0\\\hline
     $\chi_2^{(4)}$&4&4&0&$-4$&$-4$&0&$-2$&$-2$&2&2&0&0&0&0&0&1&1&$-1$&$-1$
      &0&0&0&0\\\hline
     $\chi^{(8)}$&8&8&0&$-8$&$-8$&0&0&0&0&0&0&0&0&0&0&$-1$&$-1$&1&1
      &0&0&0&0\\\hline
     $\chi^{(6)}$&6&6&$-2$&6&6&0&0&0&0&0&2&$-2$&$-2$&$-2$&$-2$&0&0&0&0
      &0&0&0&0\\\hline
     $\chi_{\alpha}^{(3)}$&3&3&-1&3&3&1&-1&-1&-1&-1&-1&-1&-1&3&3
      &0&0&0&0&-1&-1&1&1\\\hline
     $\chi_{\beta}^{(3)}$&3&3&-1&3&3&1&-1&-1&-1&-1&-1&3&3&-1&-1
      &0&0&0&0&1&1&-1&-1\\\hline
     $\chi_{\gamma}^{(3)}$&3&3&-1&3&3&-1&1&1&1&1&-1&-1&-1&3&3
      &0&0&0&0&1&1&-1&-1\\\hline
     $\chi_{\delta}^{(3)}$&3&3&-1&3&3&-1&1&1&1&1&-1&3&3&-1&-1
      &0&0&0&0&-1&-1&1&1\\\hline \hline
     $\chi_{\alpha}^{(2)}$&2&-2&0&2&-2&0&\nsqt&\msqt&\msqt&\nsqt&0&2&-2&0&0
      &1&-1&1&-1&0&0&\msqt&\nsqt\\\hline
     $\chi_{\beta}^{(2)}$&2&-2&0&2&-2&0&\msqt&\nsqt&\nsqt&\msqt&0&2&-2&0&0
      &1&-1&1&-1&0&0&\nsqt&\msqt\\\hline
     $\chi_{\gamma}^{(2)}$&2&-2&0&-2&2&0&\nsqt&\msqt&\nsqt&\msqt&0&0&0&-2&2
      &1&-1&-1&1&\msqt&\nsqt&0&0\\\hline
     $\chi_{\delta}^{(2)}$&2&-2&0&-2&2&0&\msqt&\nsqt&\msqt&\nsqt&0&0&0&-2&2
      &1&-1&-1&1&\nsqt&\msqt&0&0\\\hline
     $\chi_{\alpha}^{(4)}$&4&-4&0&4&-4&0&0&0&0&0&0&4&-4&0&0
      &-1&1&-1&1&0&0&0&0\\\hline
     $\chi_{\beta}^{(4)}$&4&-4&0&-4&4&0&0&0&0&0&0&0&0&-4&4
      &-1&1&1&-1&0&0&0&0\\\hline
     $\chi_{\alpha}^{(6)}$&6&-6&0&6&-6&0&\nsqt&\msqt&\msqt&\nsqt&0&-2&2&0&0
      &0&0&0&0&0&0&\nsqt&\msqt\\\hline
     $\chi_{\beta}^{(6)}$&6&-6&0&6&-6&0&\msqt&\nsqt&\nsqt&\msqt&0&-2&2&0&0
      &0&0&0&0&0&0&\msqt&\nsqt\\\hline
     $\chi_{\gamma}^{(6)}$&6&-6&0&-6&6&0&\nsqt&\msqt&\nsqt&\msqt&0&0&0&2&-2
      &0&0&0&0&\nsqt&\msqt&0&0\\\hline
     $\chi_{\delta}^{(6)}$&6&-6&0&-6&6&0&\msqt&\nsqt&\msqt&\nsqt&0&0&0&2&-2
      &0&0&0&0&\msqt&\nsqt&0&0\\\hline
    \end{tabular}
   \end{table}
   \normalsize
 \\
  {\bf Acknowledgements}\\
    This work was supported by Climb-Up (Pan Deng) Project of
    Department of Science and Technology in China, Chinese
    National Science Foundation and Doctoral Programme Foundation
    of Institution of Higher Education in China.
    One of the authors JD is grateful to Dr. L-G. Jin for his advice on this paper.
  
 \end{document}